\documentclass[journal=jpclcd,manuscript=letter,layout=twocolumn]{achemso}

\usepackage{braket}
\usepackage{txfonts}
\usepackage{bm}
\usepackage{color}
\definecolor{myblue}{rgb}{0,0,1}
\usepackage[breaklinks=true,colorlinks=true,linkcolor=myblue,urlcolor=myblue,citecolor=myblue]{hyperref}

\makeatletter
\let\l@addto@macro\relax
\makeatother
\usepackage[fontsize=11pt]{scrextend}

\title{Optically Addressing Circularly-Polarized Vibrations\\in Molecules}

\author{Chientzu Lin}
\author{Connor K. Terry Weatherly}
\author{Roel Tempelaar}
\email{roel.tempelaar@northwestern.edu}
\affiliation{Department of Chemistry, Northwestern University, 2145 Sheridan Road, Evanston, Illinois 60208, USA}

\begin{document}

\begin{tocentry}
\vspace{4.3\baselineskip}
\includegraphics[width=\textwidth]{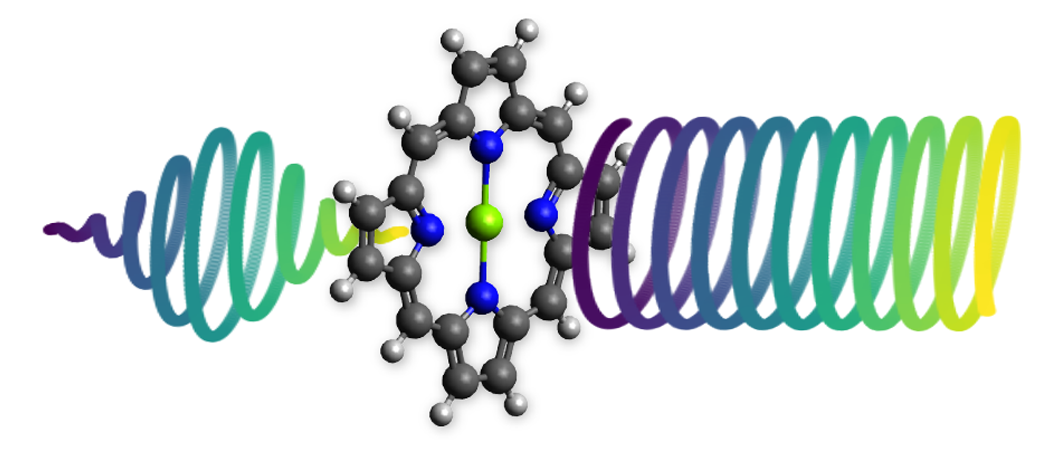}
\end{tocentry}

\begin{abstract}
Circularly-polarized (CP) vibrations are the nuclear-motion analog of CP light, emerging in molecules with non-Abelian point-group symmetry that support orthogonal and degenerate vibrational normal modes. Here, we explore the optical addressability of CP vibrations, motivated by their potential as nanoscale angular momentum states suitable for information storage and manipulation. We investigate how symmetry-breaking chemical modifications affect a molecule's capacity to support CP vibrations. Notably, we find cases where modes retain their orthogonality and degeneracy in spite of such modifications, broadening the opportunities afforded by CP vibrations beyond non-Abelian point-group molecules. Prospects for experimental implementations using CP pump–probe techniques are briefly discussed.
\end{abstract}

\maketitle

Rapid advances made in the realm of quantum information science (QIS) promote the interfacing of information theory with physics, materials science, and chemistry \cite{Scholes2025}. A fundamental question defining this interface is to what extent fundamental degrees of freedom (DOF) of light and matter can serve as carriers of information. Additional considerations include the duration for which information can be preserved and manipulated, and the means by which information can be relayed between different DOF \cite{DiVincenzo2000}. Electron spins are envisioned as canonical information carriers, with decoherence times ranging up to milliseconds at room temperature in nitrogen-vacancy centers \cite{Balasubramanian2009, Bar-Gill2013}. Photon spins, on the other hand, are leading candidates for efficient transport of information \cite{Northup2014}, due to their unmatched velocity and weak interaction strength. The manifestation of photon spin as circular optical polarization enables the transduction of such information to matter-based DOF through chiroptical interactions \cite{Salij2021}.

In some cases, electronic polarizations provide an attractive complement to spin states for information storage and manipulation. Such polarizations are commonly associated with topological states \cite{Liu2019}, at times involving a well-defined electronic angular momentum. For example, excitons in monolayer transition-metal dichalcogenides form at the inequivalent corners of the hexagonal Brillouin zone, and are characterized by electronic angular momentum, interlocked with electronic spin states for protection against rapid electronic dephasing \cite{Mak2012}. Interestingly, such excitons have been shown \cite{Delhomme2020} to exchange angular momentum with chiral phonons \cite{Zhang2014, Zhu2018}, suggesting nuclear polarizations to offer yet another avenue for information storage and manipulation.

While solids are oftentimes deemed preferred platforms for information processing, storage, and transduction, there is an increasing interest in exploring similar mechanisms within molecules, which would allow information engineering opportunities to be expanded through established molecular synthesis routes \cite{Wasielewski2020, Coronado2020}. The utilization of molecular electronic spin states as information carriers has recently been experimentally demonstrated \cite{Bayliss2020}. The possibility of addressing molecular electronic polarizations, on the other hand, has been theoretically explored in the form of ring currents in Mg-porphyrin \cite{Barth2006b, Barth2006, Barth2007, Rodriguez2012}, yet such principles are yet to be experimentally realized. In the present study, we wish to further examine the palette of DOF offered by molecules for use in information applications by considering the possibilities afforded by nuclear polarizations. Molecular vibrations can reach lifetimes of up to $\sim$140 ps at room temperature \cite{Myers1999}, which is sufficient for optical manipulation and detection. Moreover, they are highly tunable through synthetic means or through changes in chemical environment, offering a high degree of controllability.

In this Letter, we explore the scenario arising for molecules possessing non-Abelian point group symmetry, which feature orthogonal and degenerate normal modes of the nuclear structure. Such modes can be superimposed with a $\pi/2$ phase difference, yielding circularly-polarized (CP) vibrations, analogous to CP light. Moreover, CP light may serve to address such vibrations, provided that the constituent modes have allowed transition dipoles. We further investigate how symmetry-breaking chemical modifications of the molecule disrupt CP vibrations through a breaking of mode degeneracy and orthogonality. Notably, we find certain instances where degeneracies and orthogonalities survive such modifications, which opens the possibility of optically addressing CP vibrations in molecules beyond the non-Abelian point group. We envision CP pump--probe techniques as an experimental implementation of our proposed principles, which would present a stepping stone towards QIS applications.

\begin{figure}[b!]
\includegraphics
{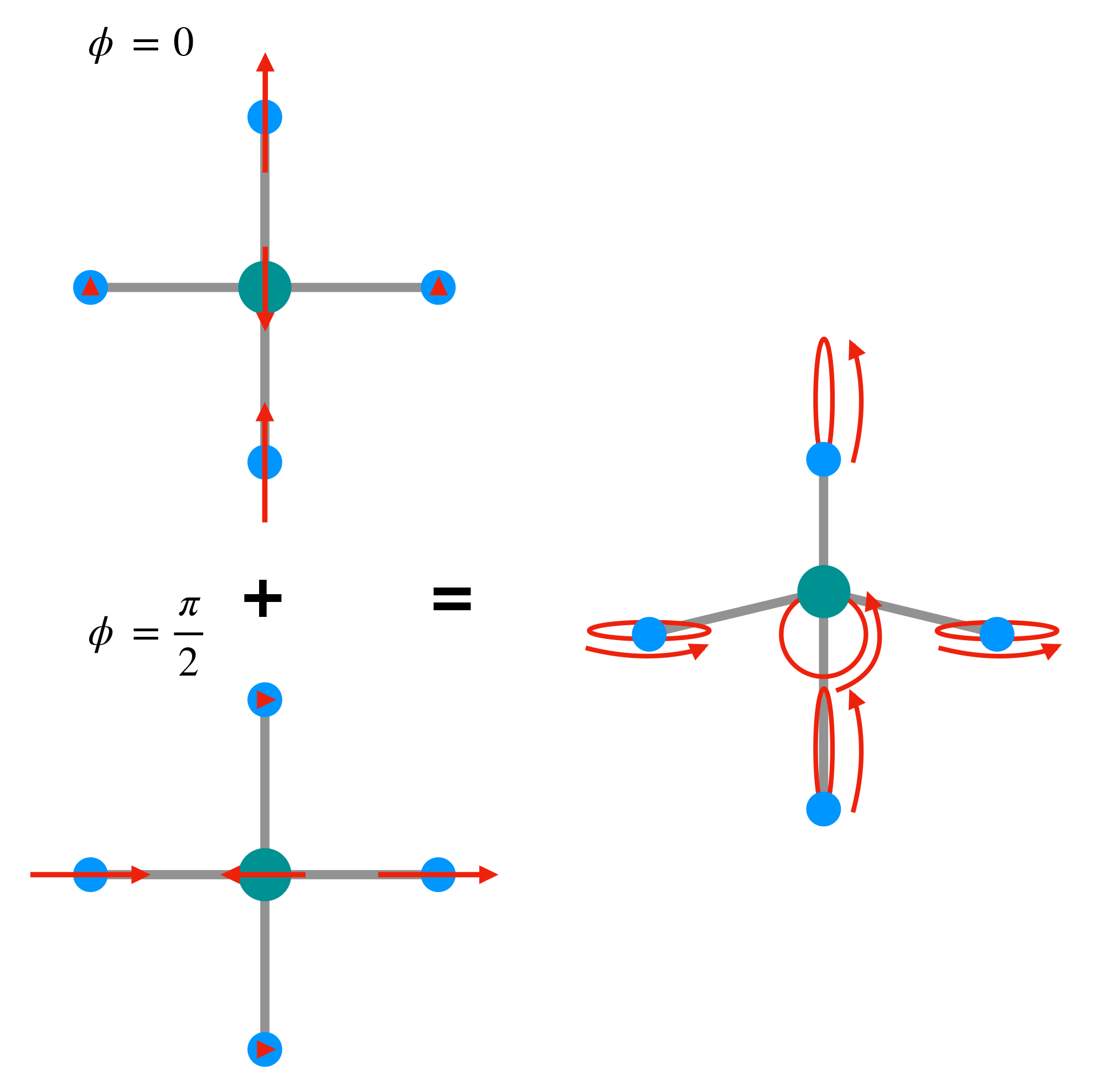}\caption{Schematic depiction of how two degenerate and orthogonal molecular vibrational normal modes (left) can be superimposed with a $\pi/2$ phase difference in order to produce a circular polarization (right). Shown are the stretching modes of xenon tetrafluoride (XeF$_4$), with the teal and blue circles depicting the Xe and F atoms, respectively, and with grey lines depicting bonds. The red lines represent the atomic path while the red arrows indicate the relative direction of movement.}
\label{fig:def}
\end{figure}

\textbf{Circularly-Polarized Vibrations.} In order to demonstrate the principles governing CP vibrations, we first consider xenon tetrafluoride (XeF$_4$), which is a planar molecule that falls within the $D_{4h}$ non-Abelian point group, implying that it possesses a 4-fold rotation axis orthogonal to the molecular plane. The infrared spectrum of xenon tetrafluoride is known to feature a peak at 586~cm$^{-1}$ assigned to a pair of degenerate and orthogonal in-plane stretching modes \cite{Claassen1963} enabled by the aforementioned 4-fold rotation axis. We performed density functional theory (DFT) calculations using Q-Chem 5.3 \cite{Shao2015} with the B3LYP functional \cite{Becke1993} and 3-21G* basis set to optimize the geometry of xenon tetrafluoride and to reproduce its stretching modes. In doing so, we found an approximate frequency of 577 cm$^{-1}$ and confirmed that the polarization directions of the modes are orthogonal. The nuclear motion associated with the modes is schematically depicted in Fig.~\ref{fig:def}. Here, it is also demonstrated how the modes can be superimposed with a $\pi/2$ phase difference in order to produce a CP vibration. This vibration is associated with internal angular momenta due to the constituent atoms \cite{Shaffer1944}, where it is noted that the total angular momentum of the molecule remains zero.

The stretching modes in xenon tetrafluoride can be optically addressed by virtue of their finite transition dipole moments (TDMs). Importantly, the transition dipole vectors align according to the mode polarizations, and are therefore orthogonal. CP light, which involves orthogonal optical field components that oscillate under a $\pi/2$ phase difference, can therefore serve to excite the CP vibration through coupling to the TDMs, provided the light it is incident in the direction normal to the XeF$_4$ plane. Once excited, this vibration will sustain after the optical field has disappeared, to the extent allowed by the vibrational lifetime. Notably, the CP vibration involves angular motion of an effective vibrational transition dipole, which in turn induces CP light emission. The resultant emitted signal not only serves to probe the CP vibrational mode, but it also represents a meaningful figure-of-merit of the degree at which this mode is sustained in time.

\begin{figure}[t!]
\includegraphics
{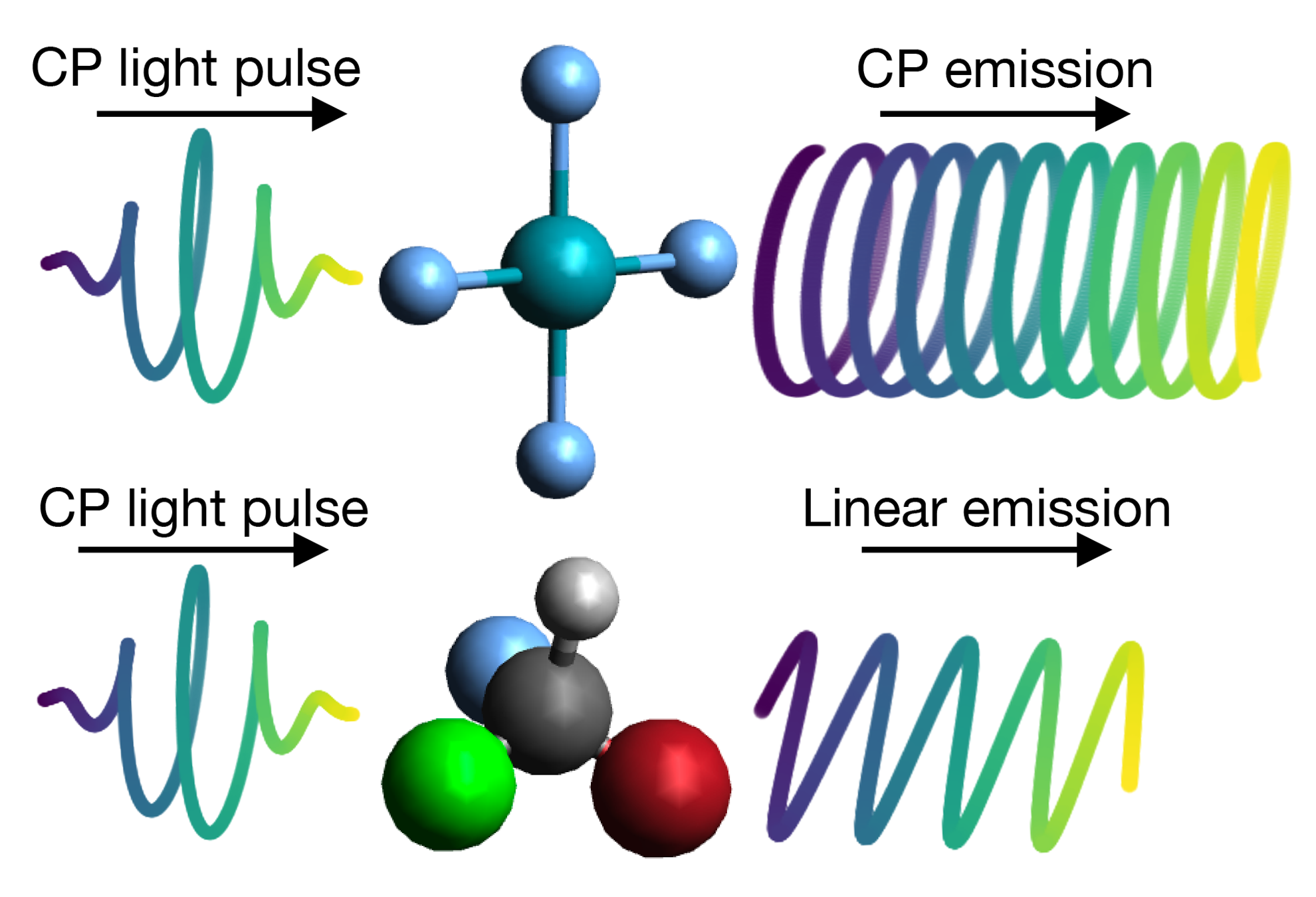}
\caption{Schematic depiction of the ability and inability of optically addressing circularly-polarized (CP) vibrations for molecules having and lacking non-Abelian point group symmetry, respectively. Shown are the examples of xenon tetrafluoride and bromochlorofluoromethane, which possess $D_{4h}$ and $C_{1}$ point-group symmetries, respectively. For the latter, emitted light is (practically) linearly-polarized, regardless of the optical excitation field polarization.}
\label{fig:compare}
\end{figure}

It is insightful to contrast the scenario of optically addressing circular polarizations of the stretching modes in xenon tetrafluoride with what is expected for molecules lacking orthogonality and degeneracy of the vibrational modes, by looking beyond the non-Abelian point groups. Perhaps the most extreme example is provided by bromochlorofluoromethane (CHBrClF), which possesses $C_1$ point-group symmetry, meaning that it only transforms onto itself through the identity operation. As a result, the molecule lacks orthogonal and degenerate vibrational modes, precluding the possibility of hosting CP vibrations. Different from the CP light emission by xenon tetrafluoride, signals emitted by bromochlorofluoromethane will therefore be (practically) linearly-polarized, regardless of the polarization of the excitation field, as is illustrated in Fig.~\ref{fig:compare}.

In order to explore the principles of optically addressing CP vibrations more quantitatively, we conducted classical simulations of the xenon tetrafluoride example. Accordingly, the normal modes of the molecule are represented by classical harmonic oscillators, which are coupled to a classical optical field taking the form of an excitation pulse. Such is governed by the Hamiltonian function
\begin{equation}
    H = \frac{1}{2}\sum_{i}\left(p_i^2+\omega_i^2 q_i^2\right)+\sum_{i}\boldsymbol{E}_{\mathrm{pulse}} \cdot \boldsymbol{\mu}_{i} \; q_i,
    \label{eq:ham}
\end{equation}
where $i$ labels the normal mode, and $\omega_i$ and $\boldsymbol{\mu}_i$ denote the corresponding frequency and transition dipole vector, respectively. Furthermore, $q_i$ and $p_i$ denote the mass-weighted position and momentum coordinates representing the normal mode harmonic oscillator. The vibrational dynamics follows from the Hamilton equations, yielding
\begin{equation}
    \frac{dq_i}{dt} = p_i, \quad
    \frac{dp_i}{dt} = - \omega_i^2 q_i-\boldsymbol{E}_\mathrm{pulse} \cdot \boldsymbol{\mu}_{i}.
    \label{eq:cl_dyn}
\end{equation}

In Eqs.~\ref{eq:ham} and \ref{eq:cl_dyn}, $\boldsymbol{E}_\mathrm{pulse}$ represents the excitation pulse, which under circular polarization is taken as the linear combination of two orthogonal electric fields with phase difference $\pi/2$. Accordingly, clockwise circular polarization is given by
\begin{equation}
    \boldsymbol{E}_{\mathrm{pulse}} = E_0 \boldsymbol{e}_x + E_{\pi/2}\boldsymbol{e}_y, \label{eq_pulse}
\end{equation}
where $\boldsymbol{e}_x$ and $\boldsymbol{e}_y$ represent the unit vectors in the $x$ and $y$ directions, taken to span the plane of the molecule. The electric fields are derived from the vector potential as
\begin{equation}
    E_\phi =- \frac{\mathrm{d}}{\mathrm{d}t} A_\phi,
\end{equation}
with the vector potential given by
\begin{equation}
    A_\phi =e^{-(t-T)^2/\sigma^2} \sin(\Omega t -\phi).
\end{equation}
Here, $T$ and $\sigma$ denote the time and temporal width of the excitation pulse, and $\Omega$ represents its angular frequency. The phase is generically denoted by $\phi$.

\begin{figure}[b!]
\includegraphics
{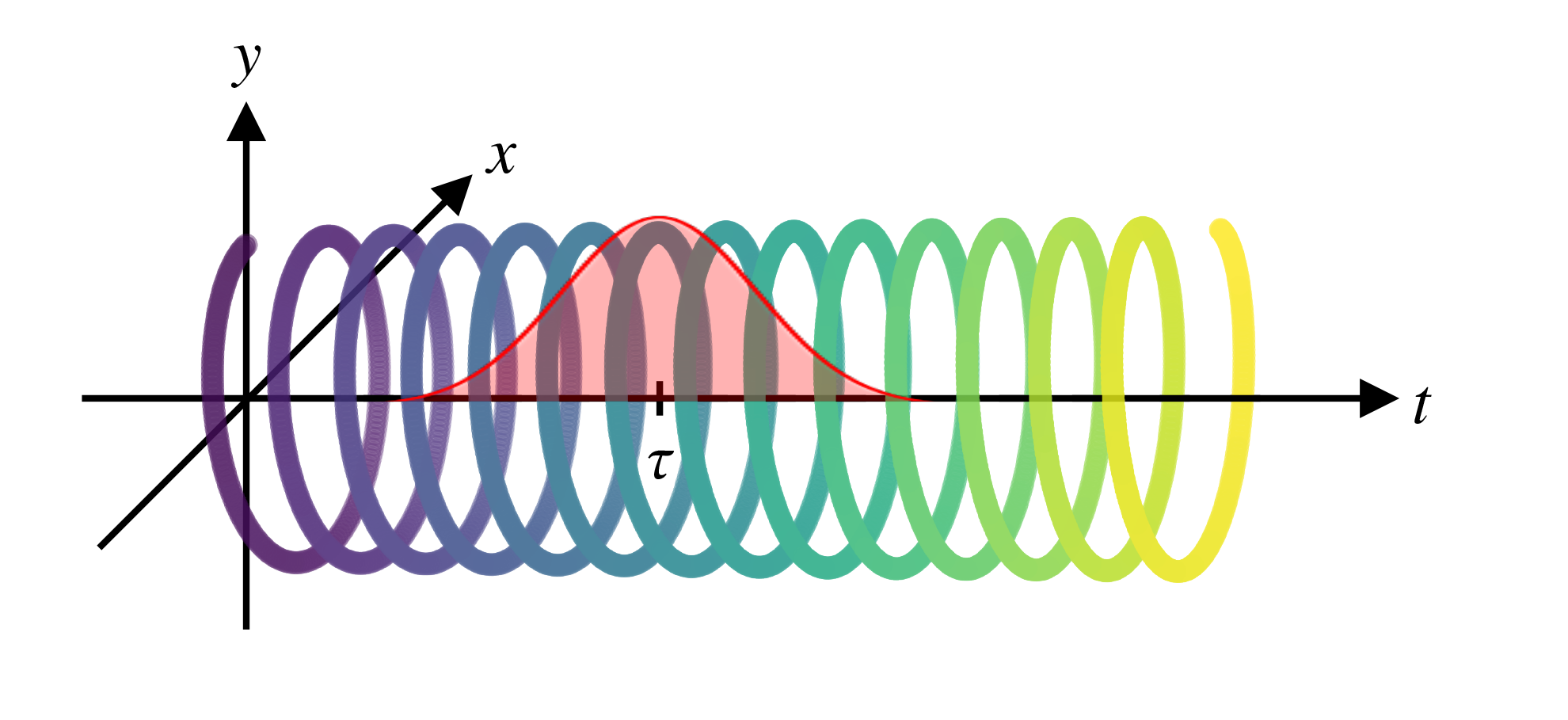}
\caption{Schematic depiction of the windowing of CP optical emission by a Gaussian function centered at time $\tau$, as applied within the short-time Fourier transform (STFT).}
\label{fig:STFT}
\end{figure}

To examine to what extent CP vibrations in xenon tetrafluoride are sustained, we examine light emission in the far-field limit, for which the signal is proportional to the oscillating transition dipoles, i.e.,
\begin{equation}
    \boldsymbol{E}_{\mathrm{signal}} \propto \sum_{i}\boldsymbol{\mu}_{i} q_i.
\end{equation}
The transient evolution of the degree of circular polarization of $\boldsymbol{E}_{\mathrm{signal}}$ then informs on the conservation of vibrational circular polarization by the nuclear structure of the molecule. In order to quantify this transient circular polarization, we project the $x$ and $y$ field components onto the complex plane, upon which we perform a short-time Fourier transform (STFT), following
\begin{eqnarray}
    \mathrm{STFT} \{ \boldsymbol{E}_{\mathrm{signal}} \} &=& \int_{-\infty}^{\infty} \left({\boldsymbol{E}}_{\mathrm{signal}} \cdot \boldsymbol{e}_x + i {\boldsymbol{E}}_{\mathrm{signal}} \cdot \boldsymbol{e}_y \right)\nonumber\\ & \cdot & W_\tau \,e^{-i \omega t} \,\mathrm{d}t.
\end{eqnarray}
An STFT signal at positive angular frequency $\omega$ then represents a circular polarization state indicative of clockwise motion of the effective vibrational transition dipole, while counter-clockwise motion yield an STFT signal at negative $\omega$. The STFT features a window function $W_\tau$, taken to be a Gaussian of the form
\begin{equation}
    W_\tau = e^{-(t-\tau)^2/\sigma^2},
\end{equation}
where $\tau$ is the instant at which the signal is being evaluated. Such windowing is necessary to resolve the continuously-oscillating optical field in time, as illustrated in Fig.~\ref{fig:STFT}, while it inevitably compromises the frequency resolution by virtue of the uncertainty principle.

\begin{figure}[t!]
\includegraphics
{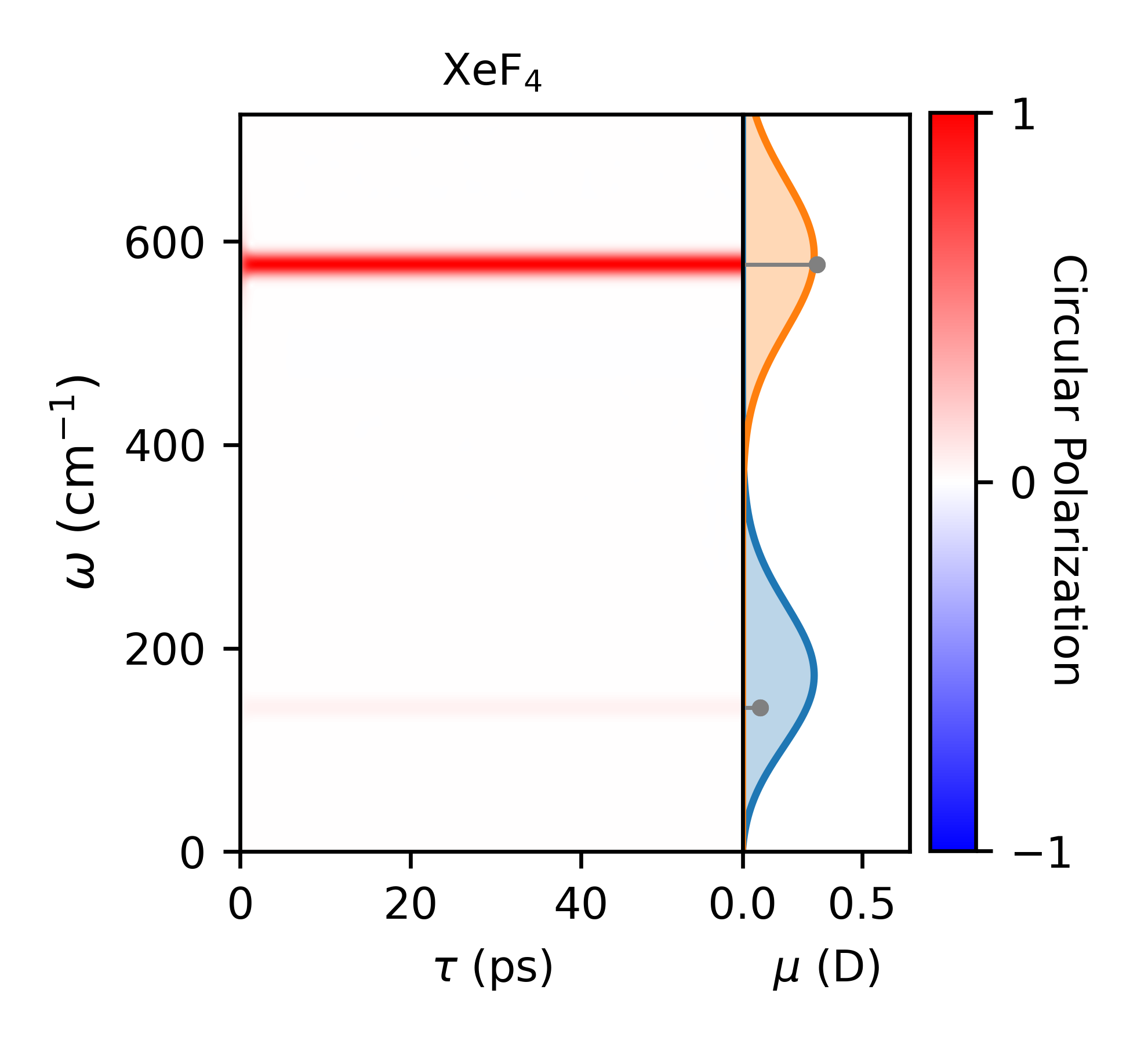}
\caption{Calculated transient circular polarization of the signal emitted by xenon tetrafluoride after being excited by a clockwise CP optical pulse. Shown on the right are two excitation pulse spectra overlaid by a stick spectrum of the TDMs (in Debye) of the normal modes in xenon tetrafluoride projected into the $xy$ plane. Shown on the left is a concatenated plot of the STFT characterization of the emitted signal due to both excitation pulses, where positive and negative circular polarization values are representative of clockwise and counter-clockwise motion of the effective vibrational transition dipole, respectively (see text for details).}
\label{fig:XeF4}
\end{figure}

In Fig.~\ref{fig:XeF4}, we show the spectrally-resolved STFT of $\boldsymbol{E}_{\mathrm{signal}}$ as a function of the time $\tau$ after xenon tetrafluoride is prepared in a clockwise CP vibrational mode. In depicting the STFT, we have taken the difference between the absolute signal intensity associated with positive and negative $\omega$ values, and mapped these onto the positive frequency domain, such that a net positive signal indicates that the effective vibrational transition dipole undergoes motion that is predominantly clockwise. Included in Fig.~\ref{fig:XeF4} is a stick spectrum representing the vibrational TDMs of xenon tetrafluoride projected onto the $xy$ plane, as obtained through the aforementioned DFT calculations. In addition to the two-fold degenerate stretching modes at 577 cm$^{-1}$, this stick spectrum shows a feature at 142 cm$^{-1}$ representative of a two-fold degenerate bending mode. We have performed two separate simulations while addressing both pairs of modes using excitation pulses with different spectral profiles that are depicted alongside the stick spectrum. The STFT data shown in Fig.~\ref{fig:XeF4} is a concatenation of the two simulations, which exhibits both pairs of modes as persistent positive features, each being broadened as a result of the aforementioned windowing. Hence, each feature represents a CP emission signal indicative of sustained clockwise vibrational circular polarizations.

\begin{figure}[t!]
\center
\includegraphics
{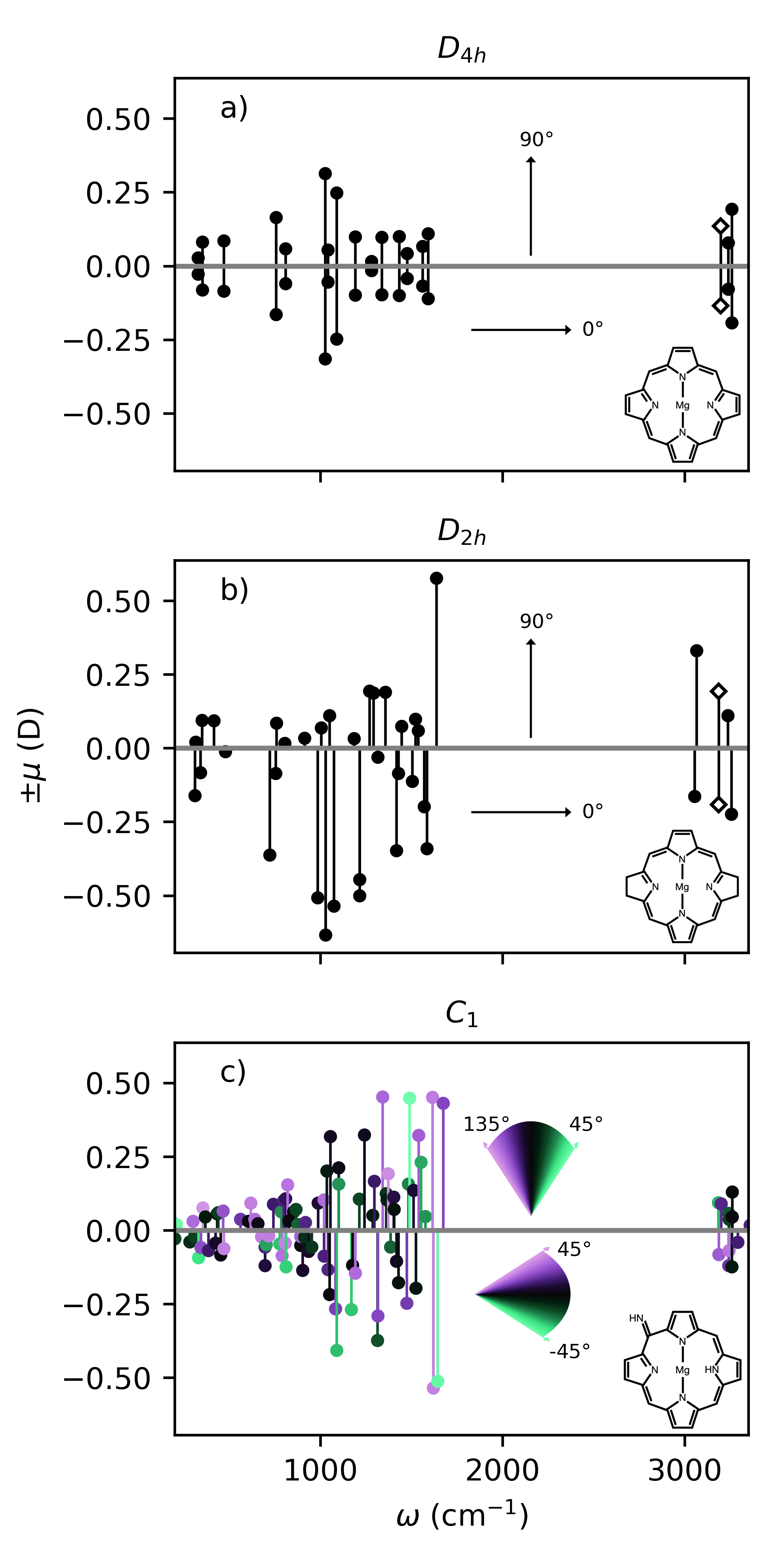}
\caption{Stick spectrum of the vibrational TDMs of Mg-porphyrin (a) which possesses $D_{4h}$ point-group symmetry. Also shown are results for the Mg-porphyrin-derivatives Mg-bacteriochlorin (b) and Mg;23H-porphyrin-22,24-diid-5-imine (c), which possess $D_{2h}$ and $C_1$ point-group symmetries, respectively. Transition dipoles oriented along the $x$ and $y$ axes are plotted in the negative and positive domains, respectively, and deviations from those axes is indicated by a color scale. Molecular geometries are shown as an inset. Open symbols denote the degenerate and orthogonal mode pair examined in Fig.~\ref{fig:new}.}
\label{fig:porphs}
\end{figure}

\textbf{Effects of Symmetry Breaking.} Thus far, we showed conserved CP vibrations in xenon tetrafluoride, which falls within the $D_{4h}$ non-Abelian point group. While this molecule serves as a useful minimal example case for vibrational circular polarizations, it serves poorly in demonstrating the overall robustness of such polarizations to symmetry-breaking chemical modifications, as its minimal size will render any modification to be relatively invasive. We therefore proceed to consider Mg-porphyrin, which is a planar molecule sharing the same point group as xenon tetrafluoride, but whose larger size allows for modifications that are comparatively more subtle.

The molecular geometry of Mg-porphyrin is depicted in Fig.~\ref{fig:porphs} (a), where we also characterize the in-plane vibrational transitions by means of a TDM stick spectrum separated into a positive and negative domain depending on whether the transition dipole vectors are oriented in the $y$ or $x$ direction, respectively. The transition dipoles shown were calculated within DFT using the B3LYP functional \cite{Becke1993} and the 6-31G* basis set. Mg-porphyrin is a much larger molecule than xenon tetrafluoride, as a result of which we now find a plethora of in-plane normal modes, some of which are concentrated in narrow frequency regions. As expected, these modes are seen to organize into degenerate and orthogonal pairs, as dictated by the $D_{4h}$ point group symmetry. We note that the mode pairs were randomly oriented in the DFT output files, and that slight degeneracy breaking occurred due to numerical errors. For the data shown in Fig.~\ref{fig:porphs}, orthogonal modes within $0.6$ cm$^{-1}$ were considered degenerate, and considering that the overall orientation of degenerate and orthogonal modes is arbitrary, they were rotated so as to align with the $x$ and $y$ directions.

\begin{figure*}
\center
\includegraphics{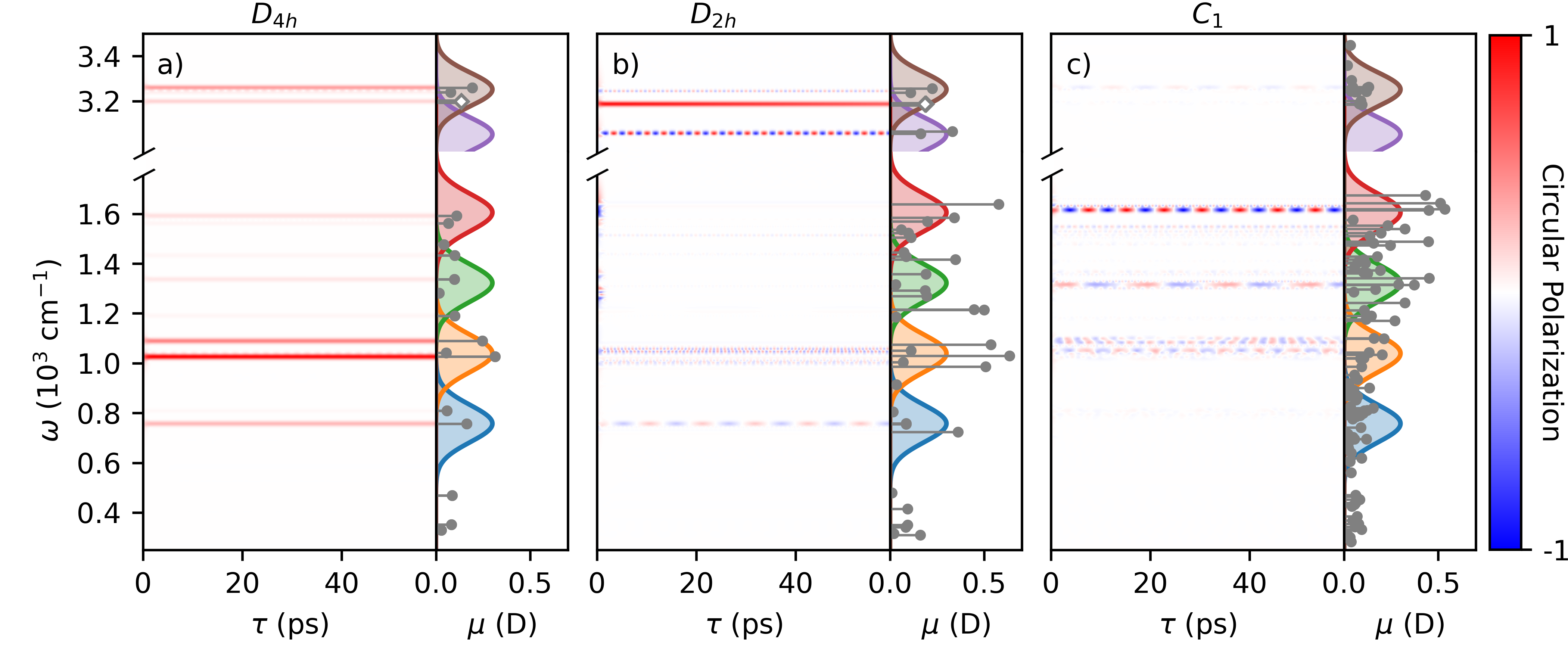}
\caption{Same as Fig.~\ref{fig:XeF4}, but for Mg-porphyrin (a) and the Mg-porphyrin-derivatives Mg-bacteriochlorin (b) and Mg;23H-porphyrin-22,24-diid-5-imine (c).}
\label{fig:all}
\end{figure*}

Based on the DFT calculated normal modes of Mg-porphyrin shown in Fig.~\ref{fig:porphs} (a), we conducted an evaluation of the transient circular polarization of $\boldsymbol{E}_\mathrm{signal}$ using the STFT analysis previously applied to xenon tetrafluoride. The results are presented in Fig.~\ref{fig:all} (a), and depicted in a similar fashion as the xenon tetrafluoride data in Fig.~\ref{fig:XeF4}. This time, however, the data is shown as a concatenation of 6 simulations under different excitation pulse spectra, necessary to cover an extended range of relevant modes. As expected, sustained CP vibrations are found for all modes, similarly to what was found for xenon tetrafluoride.

We now proceed to evaluate the effect of symmetry-breaking chemical modifications. As a first example of such a modification, we consider the addition of two hydrogens to each of the two opposing outer pyrrole cores of Mg-porphyrin. This produces Mg-bacteriochlorin, which possesses $D_{2h}$ point-group symmetry, and whose molecular geometry is shown in Fig.~\ref{fig:porphs} (b). Here, we also depict the orientation-resolved vibrational TDM stick spectrum, obtained at the same level of DFT as for Mg-porphyrin, where it can be seen that the $D_{2h}$ point-group symmetry retains an $x$ or $y$ polarization of the transition dipoles, corresponding to an alignment with the ``short'' and ``long'' molecular axes. Importantly, however, mode degeneracies have become distorted upon the applied chemical modification, concomitant with a redistribution of TDM. Compared to Mg-porphyrin, we also observe an increase in the number of modes with finite TDM, due to the reduced symmetry of Mg-bacteriochlorin. Interestingly, one mode pair has roughly retained its degeneracy, manifesting as two orthogonal features at $\sim$3190 cm$^{-1}$ with equal TDMs.

The STFT analysis of Mg-bacteriochlorin is shown in Fig.~\ref{fig:all} (b). Here, a variety of modes are seen to contribute negligibly to the STFT signal in spite of a considerable TDM, due to the absence of a complementary mode that is orthogonal and degenerate. Interestingly, certain features are seen to oscillate between positive and negative values, which are attributable to orthogonal mode pairs for which degeneracy is broken only moderately. In such case, the excitation pulse governed by Eq.~\ref{eq_pulse} will drive the associated vibrational transition dipoles under a $\pi/2$ phase difference, temporarily producing a CP vibration, but the lack of degeneracy will cause the orthogonal components to oscillate at different frequencies, inducing a drift of the phase difference. As a result, there are instances where the phase difference is inverted, producing temporal inversions of circular vibrational motion, which explains the sign alternations seen for the STFT features. An exception, however, is found for the normal mode pair at $\sim$3190 cm$^{-1}$, which is seen to exhibit sustained circular motion due to a retained degeneracy.

\begin{figure}[b!]
\center
\includegraphics
{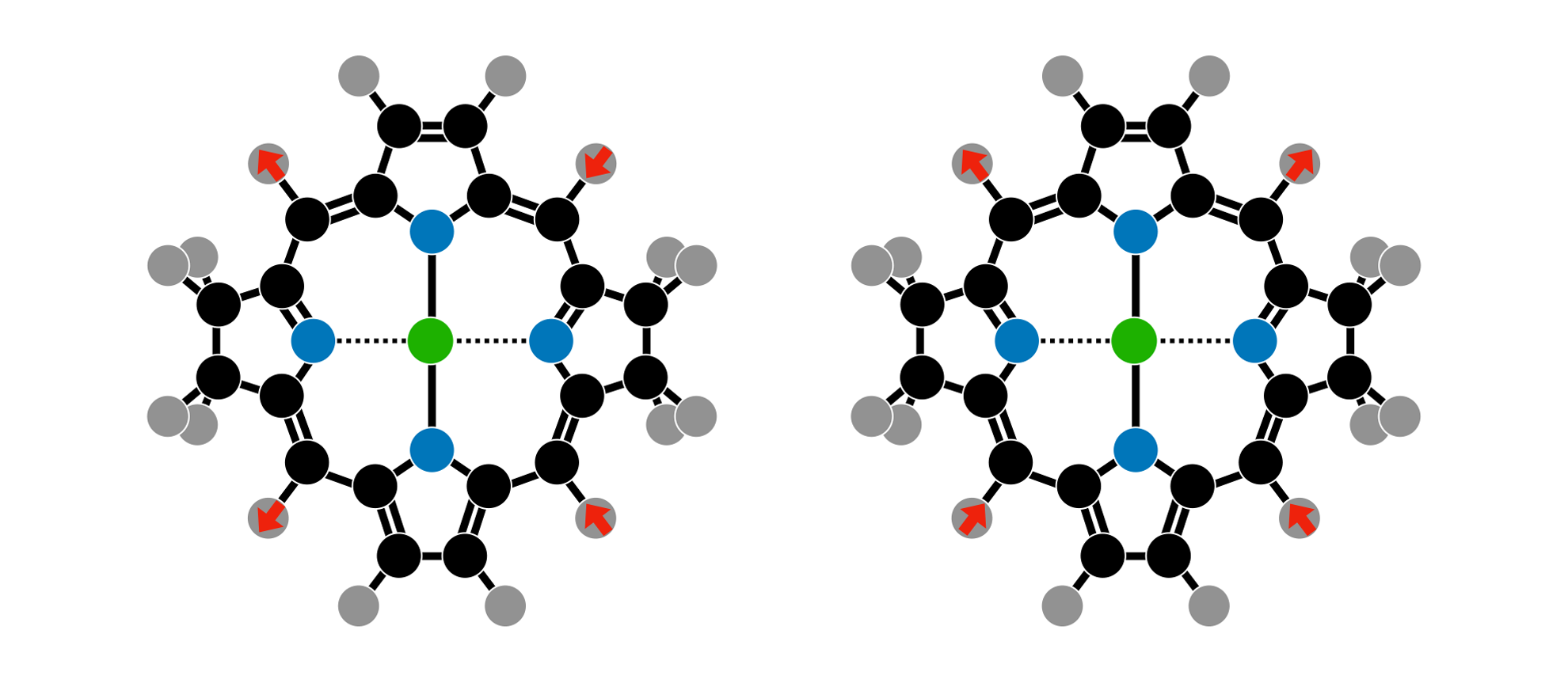}
\caption{Schematic depiction of the pair of degenerate and orthogonal vibrational modes producing a CP vibration in Mg-bacteriochlorin. This mode pair, which is also present in Mg-porphyrin, is indicated with open symbols in the vibrational TDM stick spectrum shown in Fig.~\ref{fig:all}.}
\label{fig:new}
\end{figure}

In order to analyze the $\sim$3190 cm$^{-1}$ mode pair producing a CP vibration in Mg-bacteriochlorin, we show in Fig.~\ref{fig:new} the corresponding orthogonal components. These components are seen to exclusively engage peripheral C--H stretches, combined into orthogonal combinations. These combinations turn out to be common to both Mg-bacteriochlorin and Mg-porphyrin, and it is the peripheral nature that explains their relative insensitivity to the applied chemical modification differentiating the two compounds (the combinations are indicated with open symbols in the TDM stick spectra shown in Fig.~\ref{fig:porphs}).

As a second example of a symmetry-breaking chemical modification of Mg-porphyrin, we consider the substitution of one of the peripheral hydrogens by an imine group. Specifically, this substitution is performed at one of the hydrogens engaged in the modes scrutinized in Fig.~\ref{fig:new}, yielding the compound Mg;23H-porphyrin-22,24-diid-5-imine, which shares the same $C_{1}$ point-group symmetry as bromochlorofluoromethane, and whose molecular geometry is shown in Fig.~\ref{fig:porphs} (c). Also shown in Fig.~\ref{fig:porphs} (c) is the orientation-resolved vibrational TDM stick spectrum, obtained at the same level of DFT as for Mg-porphyrin and Mg-bacteriochlorin. Here, we observe that modes are no longer oriented along the $x$ and $y$ directions, and instead show polarizations scattered across the molecular plane as a result of the lowered symmetry (deviations from $x$ and $y$ polarizations are indicated using a color scale). Upon a comparison to Mg-bacteriochlorin, the lowered symmetry is also seen to further increase the number of modes with finite TDMs. Most of these modes contribute negligibly to the STFT signal, which is depicted in Fig.~\ref{fig:all} (c). Similarly to Mg-bacteriochlorin, we find oscillating STFT features due to certain mode pairs that are within energetic proximity while having orthogonal components, yet the orthogonality should be considered accidental in this case. Notably, the modes shown in Fig.~\ref{fig:new} are absent in this case, since one of the participating hydrogens has now been substituted. Absent any orthogonal and degenerate mode pairs, no persistent CP vibrations are found in this case.

\textbf{Summary and Outlook.} In summary, we have theoretically explored CP vibrations in molecules, and the principles through which such vibrations can be optically addressed by means of CP light. We have considered two planar molecules, xenon tetrafluoride and Mg-porphyrin, each exhibiting pairs of orthogonal and degenerate normal modes of the nuclear structure that can be engaged in sustained circular motion. The presence of such mode pairs is ensured by the non-Abelian $D_{4h}$ point-group symmetry possessed by the molecules. We have furthermore shown instances where the degeneracy and orthogonality of a mode pair proves immune to a symmetry-breaking chemical modification of Mg-porphyrin, suggesting the realizability of CP vibrations in molecules beyond the non-Abelian point group.

We hope our findings offer guidance for the potential application of nuclear DOF in molecule-based QIS applications. Towards such applications, it will be of interest to extend our analysis by considering the effect of vibrational relaxation mechanisms. Such mechanisms would trivially reduce the time scale at which circular polarizations are sustained, but a quantitative analysis of the nature and magnitude of such mechanisms is nontrivial and beyond the scope of the present study. The dielectric environment of molecules is known to play an important role in the degree of vibrational relaxation, and a judicious choice of this environment (e.g., gas phase, polar and nonpolar solvents, or polymer matrix embedding) is therefore critical to optimally harnessing CP vibrations. Another optimization route is provided by temperature, the lowering of which usually suppresses vibrational relaxation \cite{Millikan1963}. It should be noted that measured room-temperature relaxation times upwards of tens of picoseconds have been reported \cite{Dlott1996, Laubereau1978, Myers1999}, suggesting that the addressing and manipulating vibrations is achievable by optical means.

With the ultimate prospect of harnessing CP vibrations for QIS applications, it will be worth investigating such vibrations by means of pump--prove experiments that invoke circular polarizations for both pump and probe pulses \cite{DalConte2015, Li2020} (at times referred to as circularly-polarized transient absorption or time-resolved circular dichroism). Accordingly, the pump pulse excites CP vibrations as detailed in this Letter, and the probe pulse would detect sustained circular motion by means of stimulated emission. A coherent two-dimensional spectroscopic extension of this experiment is also worth considering, inspired by previous applications of circular polarizations for this technique \cite{Hochstrasser2001, Cho2003}. It may be possible to performed such measurements for oriented molecules, consistent with the simulations presented in this Article. While such may be achievable through matrix embedding, an interesting alternative approach is offered by two-dimensional polymerization, such as applied to porphyrin monomers \cite{Zhong2019}. When an isotropic molecular sample is used instead, the effective magnitude of circular motion will diminish due to imperfectly-aligned molecules, but the ability to control the \emph{direction} of this motion will be retained (as can be easily verified).

It will furthermore be of interest to investigate the addressability of CP vibrations by means beyond optical excitation, through interactions with electronic ring currents \cite{Barth2006, Barth2006b, Barth2007, Rodriguez2012}, or through exchange of atomic angular momentum with chiral vibrational vibrations \cite{Abraham2024}. Altogether, such advances would help expand the range of molecular DOF amenable to QIS purposes.

\textbf{Acknowledgments.} The authors wish to thank Antonio Garz\'on-Ram\'irez for helpful discussions. This work was supported as part of the Center for Molecular Quantum Transduction, an Energy Frontier Research Center funded by the U.S.~Department of Energy, Office of Science, Basic Energy Sciences under Award \# DE-SC0021314.
 
\bibliography{bibliography}

\providecommand{\latin}[1]{#1}
\makeatletter
\providecommand{\doi}
  {\begingroup\let\do\@makeother\dospecials
  \catcode`\{=1 \catcode`\}=2 \doi@aux}
\providecommand{\doi@aux}[1]{\endgroup\texttt{#1}}
\makeatother
\providecommand*\mcitethebibliography{\thebibliography}
\csname @ifundefined\endcsname{endmcitethebibliography}  {\let\endmcitethebibliography\endthebibliography}{}
\begin{mcitethebibliography}{33}
\providecommand*\natexlab[1]{#1}
\providecommand*\mciteSetBstSublistMode[1]{}
\providecommand*\mciteSetBstMaxWidthForm[2]{}
\providecommand*\mciteBstWouldAddEndPuncttrue
  {\def\EndOfBibitem{\unskip.}}
\providecommand*\mciteBstWouldAddEndPunctfalse
  {\let\EndOfBibitem\relax}
\providecommand*\mciteSetBstMidEndSepPunct[3]{}
\providecommand*\mciteSetBstSublistLabelBeginEnd[3]{}
\providecommand*\EndOfBibitem{}
\mciteSetBstSublistMode{f}
\mciteSetBstMaxWidthForm{subitem}{(\alph{mcitesubitemcount})}
\mciteSetBstSublistLabelBeginEnd
  {\mcitemaxwidthsubitemform\space}
  {\relax}
  {\relax}

\bibitem[Scholes \latin{et~al.}(2025)Scholes, Olaya-Castro, Mukamel, Kirrander, Ni, Hedley, and Frank]{Scholes2025}
Scholes,~G.~D.; Olaya-Castro,~A.; Mukamel,~S.; Kirrander,~A.; Ni,~K.-K.; Hedley,~G.~J.; Frank,~N.~L. The Quantum Information Science Challenge for Chemistry. \emph{The Journal of Physical Chemistry Letters} \textbf{2025}, \emph{16}, 1376--1396\relax
\mciteBstWouldAddEndPuncttrue
\mciteSetBstMidEndSepPunct{\mcitedefaultmidpunct}
{\mcitedefaultendpunct}{\mcitedefaultseppunct}\relax
\EndOfBibitem
\bibitem[DiVincenzo(2000)]{DiVincenzo2000}
DiVincenzo,~D.~P. The Physical Implementation of Quantum Computation. \emph{Fortschritte der Physik} \textbf{2000}, \emph{48}, 771--783\relax
\mciteBstWouldAddEndPuncttrue
\mciteSetBstMidEndSepPunct{\mcitedefaultmidpunct}
{\mcitedefaultendpunct}{\mcitedefaultseppunct}\relax
\EndOfBibitem
\bibitem[Balasubramanian \latin{et~al.}(2009)Balasubramanian, Neumann, Twitchen, Markham, Kolesov, Mizuochi, Isoya, Achard, Beck, Tissler, Jacques, Hemmer, Jelezko, and Wrachtrup]{Balasubramanian2009}
Balasubramanian,~G.; Neumann,~P.; Twitchen,~D.; Markham,~M.; Kolesov,~R.; Mizuochi,~N.; Isoya,~J.; Achard,~J.; Beck,~J.; Tissler,~J. \latin{et~al.}  Ultralong spin coherence time in isotopically engineered diamond. \emph{Nature Materials} \textbf{2009}, \emph{8}, 383--387\relax
\mciteBstWouldAddEndPuncttrue
\mciteSetBstMidEndSepPunct{\mcitedefaultmidpunct}
{\mcitedefaultendpunct}{\mcitedefaultseppunct}\relax
\EndOfBibitem
\bibitem[Bar-Gill \latin{et~al.}(2013)Bar-Gill, Pham, Jarmola, Budker, and Walsworth]{Bar-Gill2013}
Bar-Gill,~N.; Pham,~L.~M.; Jarmola,~A.; Budker,~D.; Walsworth,~R.~L. Solid-state electronic spin coherence time approaching one second. \emph{Nature Communications} \textbf{2013}, \emph{4}, 1743\relax
\mciteBstWouldAddEndPuncttrue
\mciteSetBstMidEndSepPunct{\mcitedefaultmidpunct}
{\mcitedefaultendpunct}{\mcitedefaultseppunct}\relax
\EndOfBibitem
\bibitem[Northup and Blatt(2014)Northup, and Blatt]{Northup2014}
Northup,~T.; Blatt,~R. Quantum information transfer using photons. \emph{Nature Photonics} \textbf{2014}, \emph{8}, 356--363\relax
\mciteBstWouldAddEndPuncttrue
\mciteSetBstMidEndSepPunct{\mcitedefaultmidpunct}
{\mcitedefaultendpunct}{\mcitedefaultseppunct}\relax
\EndOfBibitem
\bibitem[Salij \latin{et~al.}(2021)Salij, Goldsmith, and Tempelaar]{Salij2021}
Salij,~A.; Goldsmith,~R.~H.; Tempelaar,~R. Theory of Apparent Circular Dichroism Reveals the Origin of Inverted and Noninverted Chiroptical Response under Sample Flipping. \emph{Journal of the American Chemical Society} \textbf{2021}, \emph{143}, 21519--21531, PMID: 34914380\relax
\mciteBstWouldAddEndPuncttrue
\mciteSetBstMidEndSepPunct{\mcitedefaultmidpunct}
{\mcitedefaultendpunct}{\mcitedefaultseppunct}\relax
\EndOfBibitem
\bibitem[Liu and Hersam(2019)Liu, and Hersam]{Liu2019}
Liu,~X.; Hersam,~M.~C. 2D materials for quantum information science. \emph{Nature Reviews Materials} \textbf{2019}, \emph{4}, 669--684\relax
\mciteBstWouldAddEndPuncttrue
\mciteSetBstMidEndSepPunct{\mcitedefaultmidpunct}
{\mcitedefaultendpunct}{\mcitedefaultseppunct}\relax
\EndOfBibitem
\bibitem[Mak \latin{et~al.}(2012)Mak, He, Shan, and Heinz]{Mak2012}
Mak,~K.~F.; He,~K.; Shan,~J.; Heinz,~T.~F. Control of valley polarization in monolayer MoS2 by optical helicity. \emph{Nature Nanotechnology} \textbf{2012}, \emph{7}, 494--498\relax
\mciteBstWouldAddEndPuncttrue
\mciteSetBstMidEndSepPunct{\mcitedefaultmidpunct}
{\mcitedefaultendpunct}{\mcitedefaultseppunct}\relax
\EndOfBibitem
\bibitem[Delhomme \latin{et~al.}(2020)Delhomme, Vaclavkova, Slobodeniuk, Orlita, Potemski, Basko, Watanabe, Taniguchi, Mauro, Barreteau, Giannini, Morpurgo, Ubrig, and Faugeras]{Delhomme2020}
Delhomme,~A.; Vaclavkova,~D.; Slobodeniuk,~A.; Orlita,~M.; Potemski,~M.; Basko,~D.~M.; Watanabe,~K.; Taniguchi,~T.; Mauro,~D.; Barreteau,~C. \latin{et~al.}  Flipping exciton angular momentum with chiral phonons in MoSe2/WSe2 heterobilayers. \emph{2D Materials} \textbf{2020}, \emph{7}, 041002\relax
\mciteBstWouldAddEndPuncttrue
\mciteSetBstMidEndSepPunct{\mcitedefaultmidpunct}
{\mcitedefaultendpunct}{\mcitedefaultseppunct}\relax
\EndOfBibitem
\bibitem[Zhang and Niu(2014)Zhang, and Niu]{Zhang2014}
Zhang,~L.; Niu,~Q. Angular Momentum of Phonons and the Einstein–de Haas Effect. \emph{Physical Review Letters} \textbf{2014}, \emph{112}, 085503\relax
\mciteBstWouldAddEndPuncttrue
\mciteSetBstMidEndSepPunct{\mcitedefaultmidpunct}
{\mcitedefaultendpunct}{\mcitedefaultseppunct}\relax
\EndOfBibitem
\bibitem[Zhu \latin{et~al.}(2018)Zhu, Yi, Li, Xiao, Zhang, Yang, Kaindl, Li, Wang, and Zhang]{Zhu2018}
Zhu,~H.; Yi,~J.; Li,~M.-Y.; Xiao,~J.; Zhang,~L.; Yang,~C.-W.; Kaindl,~R.~A.; Li,~L.-J.; Wang,~Y.; Zhang,~X. Observation of chiral phonons. \emph{Science} \textbf{2018}, \emph{359}, 579--582\relax
\mciteBstWouldAddEndPuncttrue
\mciteSetBstMidEndSepPunct{\mcitedefaultmidpunct}
{\mcitedefaultendpunct}{\mcitedefaultseppunct}\relax
\EndOfBibitem
\bibitem[Wasielewski \latin{et~al.}(2020)Wasielewski, Forbes, Frank, Kowalski, Scholes, Yuen-Zhou, Baldo, Freedman, Goldsmith, Goodson, Kirk, McCusker, Ogilvie, Shultz, Stoll, and Whaley]{Wasielewski2020}
Wasielewski,~M.~R.; Forbes,~M. D.~E.; Frank,~N.~L.; Kowalski,~K.; Scholes,~G.~D.; Yuen-Zhou,~J.; Baldo,~M.~A.; Freedman,~D.~E.; Goldsmith,~R.~H.; Goodson,~T. \latin{et~al.}  Exploiting chemistry and molecular systems for quantum information science. \emph{Nature Reviews Chemistry} \textbf{2020}, \emph{4}, 490--504\relax
\mciteBstWouldAddEndPuncttrue
\mciteSetBstMidEndSepPunct{\mcitedefaultmidpunct}
{\mcitedefaultendpunct}{\mcitedefaultseppunct}\relax
\EndOfBibitem
\bibitem[Coronado(2020)]{Coronado2020}
Coronado,~E. Molecular magnetism: from chemical design to spin control in molecules, materials and devices. \emph{Nature Reviews Materials} \textbf{2020}, \emph{5}, 87--104\relax
\mciteBstWouldAddEndPuncttrue
\mciteSetBstMidEndSepPunct{\mcitedefaultmidpunct}
{\mcitedefaultendpunct}{\mcitedefaultseppunct}\relax
\EndOfBibitem
\bibitem[Bayliss \latin{et~al.}(2020)Bayliss, Laorenza, Mintun, Kovos, Freedman, and Awschalom]{Bayliss2020}
Bayliss,~S.~L.; Laorenza,~D.~W.; Mintun,~P.~J.; Kovos,~B.~D.; Freedman,~D.~E.; Awschalom,~D.~D. Optically addressable molecular spins for quantum information processing. \emph{Science} \textbf{2020}, \emph{370}, 1309--1312\relax
\mciteBstWouldAddEndPuncttrue
\mciteSetBstMidEndSepPunct{\mcitedefaultmidpunct}
{\mcitedefaultendpunct}{\mcitedefaultseppunct}\relax
\EndOfBibitem
\bibitem[Barth and Manz(2006)Barth, and Manz]{Barth2006b}
Barth,~I.; Manz,~J. Periodic Electron Circulation Induced by Circularly Polarized Laser Pulses: Quantum Model Simulations for Mg Porphyrin. \emph{Angewandte Chemie International Edition} \textbf{2006}, \emph{45}, 2962--2965\relax
\mciteBstWouldAddEndPuncttrue
\mciteSetBstMidEndSepPunct{\mcitedefaultmidpunct}
{\mcitedefaultendpunct}{\mcitedefaultseppunct}\relax
\EndOfBibitem
\bibitem[Barth \latin{et~al.}(2006)Barth, Manz, Shigeta, and Yagi]{Barth2006}
Barth,~I.; Manz,~J.; Shigeta,~Y.; Yagi,~K. Unidirectional Electronic Ring Current Driven by a Few Cycle Circularly Polarized Laser Pulse: Quantum Model Simulations for Mg--Porphyrin. \emph{Journal of the American Chemical Society} \textbf{2006}, \emph{128}, 7043--7049\relax
\mciteBstWouldAddEndPuncttrue
\mciteSetBstMidEndSepPunct{\mcitedefaultmidpunct}
{\mcitedefaultendpunct}{\mcitedefaultseppunct}\relax
\EndOfBibitem
\bibitem[Barth and Manz(2007)Barth, and Manz]{Barth2007}
Barth,~I.; Manz,~J. Electric ring currents in atomic orbitals and magnetic fields induced by short intense circularly polarized $\ensuremath{\pi}$ laser pulses. \emph{Physical Review A} \textbf{2007}, \emph{75}, 012510\relax
\mciteBstWouldAddEndPuncttrue
\mciteSetBstMidEndSepPunct{\mcitedefaultmidpunct}
{\mcitedefaultendpunct}{\mcitedefaultseppunct}\relax
\EndOfBibitem
\bibitem[Rodriguez and Mukamel(2012)Rodriguez, and Mukamel]{Rodriguez2012}
Rodriguez,~J.~J.; Mukamel,~S. Probing Ring Currents in Mg-Porphyrins by Pump--Probe Spectroscopy. \emph{The Journal of Physical Chemistry A} \textbf{2012}, \emph{116}, 11095--11100, PMID: 22881200\relax
\mciteBstWouldAddEndPuncttrue
\mciteSetBstMidEndSepPunct{\mcitedefaultmidpunct}
{\mcitedefaultendpunct}{\mcitedefaultseppunct}\relax
\EndOfBibitem
\bibitem[Myers \latin{et~al.}(1999)Myers, Shigeiwa, Fayer, and Silbey]{Myers1999}
Myers,~D.; Shigeiwa,~M.; Fayer,~M.; Silbey,~R. Non-exponential relaxation of a single quantum vibrational excitation of a large molecule in collision free gas phase at elevated temperature. \emph{Chemical Physics Letters} \textbf{1999}, \emph{312}, 399--406\relax
\mciteBstWouldAddEndPuncttrue
\mciteSetBstMidEndSepPunct{\mcitedefaultmidpunct}
{\mcitedefaultendpunct}{\mcitedefaultseppunct}\relax
\EndOfBibitem
\bibitem[Claassen \latin{et~al.}(1963)Claassen, Chernick, and Malm]{Claassen1963}
Claassen,~H.~H.; Chernick,~C.~L.; Malm,~J.~G. Vibrational Spectra and Structure of Xenon Tetrafluoride. \emph{Journal of the American Chemical Society} \textbf{1963}, \emph{85}, 1927--1928\relax
\mciteBstWouldAddEndPuncttrue
\mciteSetBstMidEndSepPunct{\mcitedefaultmidpunct}
{\mcitedefaultendpunct}{\mcitedefaultseppunct}\relax
\EndOfBibitem
\bibitem[Shao \latin{et~al.}(2015)Shao, Gan, Epifanovsky, Gilbert, Wormit, Kussmann, Lange, Behn, Deng, Feng, Ghosh, Goldey, Horn, Jacobson, Kaliman, Khaliullin, Kuś, Landau, Liu, Proynov, Rhee, Richard, Rohrdanz, Steele, Sundstrom, III, Zimmerman, Zuev, Albrecht, Alguire, Austin, Beran, Bernard, Berquist, Brandhorst, Bravaya, Brown, Casanova, Chang, Chen, Chien, Closser, Crittenden, Diedenhofen, Jr., Do, Dutoi, Edgar, Fatehi, Fusti-Molnar, Ghysels, Golubeva-Zadorozhnaya, Gomes, Hanson-Heine, Harbach, Hauser, Hohenstein, Holden, Jagau, Ji, Kaduk, Khistyaev, Kim, Kim, King, Klunzinger, Kosenkov, Kowalczyk, Krauter, Lao, Laurent, Lawler, Levchenko, Lin, Liu, Livshits, Lochan, Luenser, Manohar, Manzer, Mao, Mardirossian, Marenich, Maurer, Mayhall, Neuscamman, Oana, Olivares-Amaya, O’Neill, Parkhill, Perrine, Peverati, Prociuk, Rehn, Rosta, Russ, Sharada, Sharma, Small, Sodt, Stein, Stück, Su, Thom, Tsuchimochi, Vanovschi, Vogt, Vydrov, Wang, Watson, Wenzel, White, Williams, Yang, Yeganeh, Yost, You, Zhang,
  Zhang, Zhao, Brooks, Chan, Chipman, Cramer, III, Gordon, Hehre, Klamt, III, Schmidt, Sherrill, Truhlar, Warshel, Xu, Aspuru-Guzik, Baer, Bell, Besley, Chai, Dreuw, Dunietz, Furlani, Gwaltney, Hsu, Jung, Kong, Lambrecht, Liang, Ochsenfeld, Rassolov, Slipchenko, Subotnik, Voorhis, Herbert, Krylov, Gill, and and]{Shao2015}
Shao,~Y.; Gan,~Z.; Epifanovsky,~E.; Gilbert,~A.~T.; Wormit,~M.; Kussmann,~J.; Lange,~A.~W.; Behn,~A.; Deng,~J.; Feng,~X. \latin{et~al.}  Advances in molecular quantum chemistry contained in the Q-Chem 4 program package. \emph{Molecular Physics} \textbf{2015}, \emph{113}, 184--215\relax
\mciteBstWouldAddEndPuncttrue
\mciteSetBstMidEndSepPunct{\mcitedefaultmidpunct}
{\mcitedefaultendpunct}{\mcitedefaultseppunct}\relax
\EndOfBibitem
\bibitem[Becke(1993)]{Becke1993}
Becke,~A.~D. Density‐functional thermochemistry. III. The role of exact exchange. \emph{The Journal of Chemical Physics} \textbf{1993}, \emph{98}, 5648--5652\relax
\mciteBstWouldAddEndPuncttrue
\mciteSetBstMidEndSepPunct{\mcitedefaultmidpunct}
{\mcitedefaultendpunct}{\mcitedefaultseppunct}\relax
\EndOfBibitem
\bibitem[Shaffer(1944)]{Shaffer1944}
Shaffer,~W.~H. Degenerate Modes of Vibration and Perturbations in Polyatomic Molecules. \emph{Reviews of Modern Physics} \textbf{1944}, \emph{16}, 245--259\relax
\mciteBstWouldAddEndPuncttrue
\mciteSetBstMidEndSepPunct{\mcitedefaultmidpunct}
{\mcitedefaultendpunct}{\mcitedefaultseppunct}\relax
\EndOfBibitem
\bibitem[Millikan and White(1963)Millikan, and White]{Millikan1963}
Millikan,~R.~C.; White,~D.~R. Systematics of Vibrational Relaxation. \emph{The Journal of Chemical Physics} \textbf{1963}, \emph{39}, 3209--3213\relax
\mciteBstWouldAddEndPuncttrue
\mciteSetBstMidEndSepPunct{\mcitedefaultmidpunct}
{\mcitedefaultendpunct}{\mcitedefaultseppunct}\relax
\EndOfBibitem
\bibitem[Dlott \latin{et~al.}(1996)Dlott, Fayer, Hill, Rella, Suslick, and Ziegler]{Dlott1996}
Dlott,~D.~D.; Fayer,~M.~D.; Hill,~J.~R.; Rella,~C.~W.; Suslick,~K.~S.; Ziegler,~C.~J. Vibrational Relaxation in Metalloporphyrin CO Complexes. \emph{Journal of the American Chemical Society} \textbf{1996}, \emph{118}, 7853--7854\relax
\mciteBstWouldAddEndPuncttrue
\mciteSetBstMidEndSepPunct{\mcitedefaultmidpunct}
{\mcitedefaultendpunct}{\mcitedefaultseppunct}\relax
\EndOfBibitem
\bibitem[Laubereau \latin{et~al.}(1978)Laubereau, Fischer, Spanner, and Kaiser]{Laubereau1978}
Laubereau,~A.; Fischer,~S.; Spanner,~K.; Kaiser,~W. Vibrational population lifetimes of polyatomic molecules in liquids. \emph{Chemical Physics} \textbf{1978}, \emph{31}, 335--344\relax
\mciteBstWouldAddEndPuncttrue
\mciteSetBstMidEndSepPunct{\mcitedefaultmidpunct}
{\mcitedefaultendpunct}{\mcitedefaultseppunct}\relax
\EndOfBibitem
\bibitem[Dal~Conte \latin{et~al.}(2015)Dal~Conte, Bottegoni, Pogna, De~Fazio, Ambrogio, Bargigia, D'Andrea, Lombardo, Bruna, Ciccacci, Ferrari, Cerullo, and Finazzi]{DalConte2015}
Dal~Conte,~S.; Bottegoni,~F.; Pogna,~E. A.~A.; De~Fazio,~D.; Ambrogio,~S.; Bargigia,~I.; D'Andrea,~C.; Lombardo,~A.; Bruna,~M.; Ciccacci,~F. \latin{et~al.}  Ultrafast valley relaxation dynamics in monolayer ${\mathrm{MoS}}_{2}$ probed by nonequilibrium optical techniques. \emph{Physical Review B} \textbf{2015}, \emph{92}, 235425\relax
\mciteBstWouldAddEndPuncttrue
\mciteSetBstMidEndSepPunct{\mcitedefaultmidpunct}
{\mcitedefaultendpunct}{\mcitedefaultseppunct}\relax
\EndOfBibitem
\bibitem[Li \latin{et~al.}(2020)Li, He, Luo, Lu, and Wu]{Li2020}
Li,~Y.; He,~S.; Luo,~X.; Lu,~X.; Wu,~K. Strong Spin-Selective Optical Stark Effect in Lead Halide Perovskite Quantum Dots. \emph{The Journal of Physical Chemistry Letters} \textbf{2020}, \emph{11}, 3594--3600, PMID: 32310664\relax
\mciteBstWouldAddEndPuncttrue
\mciteSetBstMidEndSepPunct{\mcitedefaultmidpunct}
{\mcitedefaultendpunct}{\mcitedefaultseppunct}\relax
\EndOfBibitem
\bibitem[Hochstrasser(2001)]{Hochstrasser2001}
Hochstrasser,~R.~M. Two-dimensional IR-spectroscopy: polarization anisotropy effects. \emph{Chemical Physics} \textbf{2001}, \emph{266}, 273--284\relax
\mciteBstWouldAddEndPuncttrue
\mciteSetBstMidEndSepPunct{\mcitedefaultmidpunct}
{\mcitedefaultendpunct}{\mcitedefaultseppunct}\relax
\EndOfBibitem
\bibitem[Cho(2003)]{Cho2003}
Cho,~M. Two-dimensional circularly polarized pump–probe spectroscopy. \emph{The Journal of Chemical Physics} \textbf{2003}, \emph{119}, 7003--7016\relax
\mciteBstWouldAddEndPuncttrue
\mciteSetBstMidEndSepPunct{\mcitedefaultmidpunct}
{\mcitedefaultendpunct}{\mcitedefaultseppunct}\relax
\EndOfBibitem
\bibitem[Zhong \latin{et~al.}(2019)Zhong, Cheng, Park, Ray, Brown, Mujid, Lee, Zhou, Suh, Lee, Mannix, Kang, Sibener, Muller, and Park]{Zhong2019}
Zhong,~Y.; Cheng,~B.; Park,~C.; Ray,~A.; Brown,~S.; Mujid,~F.; Lee,~J.-U.; Zhou,~H.; Suh,~J.; Lee,~K.-H. \latin{et~al.}  Wafer-scale synthesis of monolayer two-dimensional porphyrin polymers for hybrid superlattices. \emph{Science} \textbf{2019}, \emph{366}, 1379--1384\relax
\mciteBstWouldAddEndPuncttrue
\mciteSetBstMidEndSepPunct{\mcitedefaultmidpunct}
{\mcitedefaultendpunct}{\mcitedefaultseppunct}\relax
\EndOfBibitem
\bibitem[Abraham and Nitzan(2024)Abraham, and Nitzan]{Abraham2024}
Abraham,~E.; Nitzan,~A. Quantifying the Chirality of Vibrational Modes in Helical Molecular Chains. \emph{Physical Review Letters} \textbf{2024}, \emph{133}, 268001\relax
\mciteBstWouldAddEndPuncttrue
\mciteSetBstMidEndSepPunct{\mcitedefaultmidpunct}
{\mcitedefaultendpunct}{\mcitedefaultseppunct}\relax
\EndOfBibitem
\end{mcitethebibliography}

\end{document}